\documentstyle [preprint,aps] {revtex}
\begin{document}
\draft %revtex
%\preprint{}
         %$\bigsqcup\bigsqcup$}
%%%%%%%%%%%%%
%   Title   %
%%%%%%%%%%%%%
\title{
Suppression of persistent currents
          in 1-D disordered rings
          by Coulomb interaction}
\author{H. Kato and D. Yoshioka}
\address{Institute of Physics,
         College of Arts and Sciences,
         University of Tokyo, \\Komaba, Meguro-ku, Tokyo 153, Japan}
\maketitle
%%%%%%%%%%%%%%%%
%   Abstract   %
%%%%%%%%%%%%%%%%
\begin{abstract}
\par
Effects of Coulomb interaction on persistent currents
in disordered one-dimensional rings are numerically investigated.
First of all effectiveness of the Hartree-Fock approximation is
established on small systems.
Then the calculations are done for systems
with 40 electrons in 100 sites.
It is found that the amplitude of the average persistent current
in the diffusive regime is suppressed
as the strength of the Coulomb interaction increases.
The suppression of the current is stronger
in larger rings than in smaller ones.
The enhancement of the current by the electron-electron interaction
was not observed in the diffusive regime.
\end{abstract}
%\maketitle
\narrowtext\noindent
\par
%%%%%%%%%%%%%%%%%%%%
%   Introduction   %
%%%%%%%%%%%%%%%%%%%%
About ten years ago, B\"uttiker {\em et al.} \cite{but} suggested that
equilibrium persistent current could exist
in a mesoscopic normal metal ring pierced by a magnetic flux,
if the size of the ring is so small that
the coherence of electrons are kept in the whole system.
Later the existence of the persistent current
was confirmed by three experiments.
The first experiment\cite{exp1} was done
on many isolated copper rings,
the second one\cite{exp2} on a single gold ring,
the third one\cite{exp3} on a GaAs/AlGaAs semiconductor.
In the third experiment the system is in the ballistic regime,
and the size of the persistent current is
in reasonable agreement with theories:
The current is of the order of $I_0=ev_{{\rm F}}/L$,
where $v_{{\rm F}}$ is the Fermi velocity
and $L$ is the circumference of the ring.
On the other hand,
in the former two experiments the system is in the diffusive regime.
In this case simple theories
which neglect the electron-electron interaction
had predicted the persistent current of the order of
$I \simeq I_0/M$ where $M$ is the number of transverse channels
\cite{schmid,g2,alt}.
However, the persistent current observed was greater than that.
Especially in the second experiment the magnitude of the current
was comparable to the clean case: $I \sim I_0$.
We might think that the discrepancy comes from inadequacy
of the approximations employed.
However, numerical calculations done
by Montambaux {\em et al.} \cite{mon1,mon2}
for non-interacting electrons in disordered rings
also cannot explain the large magnitude of the current
as seen in the second experiment.
Thus it is natural to consider that
the discrepancy between theoretical calculations
and experimental data is not due to the approximation employed
in the theory but due to a defect of the model
where electron-electron interaction is neglected.\par
Models with electron-electron interaction have been investigated
both analytically and numerically.
In some analytic calculations \cite{ae,es,kop}
the diagrams called cooperon
which most contribute to the flux dependence of the energy
were considered and resulted in larger current
than that in non-interacting case.
However, it was found that the effect of higher diagrams
suppress the current \cite{sa}.
In numerical calculations
effects of the electron-electron interaction were investigated
by exact diagonalization of its Hamiltonian
for small size rings.
In the case of a one-dimensional discrete-lattice ring the interaction,
which are both long ranged \cite{ab} and short ranged \cite{monc},
is found to suppress the average persistent current.
On the contrary, starting from continuum model \cite{mg2}
it was suggested that the average persistent current was enhanced
by the electron-electron interaction.
Thus, in both analytical and numerical calculations
the role of the electron-electron interaction
for the persistent current is controversial. \par
%%%%%%%%%%%%%%%%%%
%   Motivation   %
%%%%%%%%%%%%%%%%%%
The difficulty in the present problem is that
there is no guiding principle for the analytical calculation.
It is not clear which diagrams we should take into account.
On the other hand numerical investigations,
where we can obtain results free from approximations,
are limited to very small systems.
Thus it is desirable to do numerical calculation for much larger systems.
This is what we seek in this paper:
to numerically investigate the effects of the Coulomb interaction
on the persistent current in a larger sample than ever investigated
and to find whether the results will be different
from those of the small size or not.
In order to study larger systems
we use the Hartree-Fock approximation (HFA).
We first compare the results of the HFA
with those by the exact diagonalization for small systems.
We find that the HFA gives qualitatively correct results.
So we apply HFA to systems with 100 sites
for which exact diagonalization is impossible.\par
%%%%%%%%%%%%%%
%   Method   %
%%%%%%%%%%%%%%
We consider a one-dimensional disordered lattice ring
with $N_s$ sites pierced by a magnetic flux $\phi$
in which electrons mutually interact
with long-range ($1/r$) Coulomb repulsion.
For simplicity we neglect the spin degrees of freedom of electrons.
We adopt the tight-binding model for kinetic term.
The impurity potential is introduced
by random site-energy $\varepsilon_i$
which has random uniform distribution
with width $W$ $(-W/2 \leq \varepsilon_i \leq W/2)$.
Thus,
\begin{eqnarray}
\label{ex}
{\cal H}&=&-t\sum_{i=1}^{N_s}
  ({\rm e}^{{\rm i} \theta} a_{i+1}^{\dagger} a_i
        + {\rm e}^{-{\rm i} \theta} a_i^{\dagger} a_{i+1})
        + \sum_{i=1}^{N_s} \varepsilon_i a_i^{\dagger} a_i \nonumber \\
    & & + \frac12 \sum_{i \ne j} \frac{V}{N_s}
                   \frac{1}{|2\sin[\frac{\pi}{N_s}(i-j)]|}
                                  a_i^{\dagger} a_j^{\dagger} a_j a_i,
\end{eqnarray}
where $a_i$ $(a_i^{\dagger})$ is
an annihilation (creation) operator of a spinless fermion
at ${\em i}$-site.
The $N_s+1$-th site is identical to the first site.
The effect of a flux ($\phi$) piercing a ring
is introduced through the phase factor ${\rm e^{{\rm i} \theta}}$
which is gained when a electron hops to the neighboring site,
thus $\theta=2\pi \phi/N_s \phi_0$,
where $\phi_0=h/e$.
Coulomb repulsion is parameterized by $V (>0)$. \par
In the HFA ${\cal H}$ is replaced by ${\cal H_{{\rm HF}}}$,
\begin{eqnarray}
\label{hf}
{\cal H}_{{\rm HF}}&=&-t\sum_{i=1}^{N_s}
({\rm e}^{{\rm i} \theta} a_{i+1}^{\dagger}a_i
+ {\rm e}^{-{\rm i} \theta} a_i^{\dagger} a_{i+1})
 + \sum_{i=1}^{N_s} \varepsilon_i a_i^{\dagger} a_i\nonumber \\
& &+ \frac12 \sum_{i,j} \frac{V}{N_s}
\frac{(\langle a_j^{\dagger} a_j \rangle a_i^{\dagger} a_i
- \langle a_j^{\dagger} a_i \rangle a_i^{\dagger} a_j)}
{|2\sin[\frac{\pi}{N_s}(i-j)]|},
\end{eqnarray}
where $\langle \cdots \rangle$ denotes the expectation value
with respect to self-consistent HF eigenstate
$|\Psi \rangle =
b_{1}^{\dagger} b_{2}^{\dagger}
\cdots b_{{N_{e}}}^{\dagger} |0 \rangle$,
where
$b_n$ $(b_n^{\dagger})$ is
the electron-annihilation (creation) operator
of the $n$-th eigenstate of ${\cal H_{{\rm HF}}}$
and $N_e$ is the total number of electrons.
The self-consistency is achieved by iteration.
Once the Hamiltonian is solved the persistent current
is calculated by the formula,
\begin{eqnarray}
I(\phi)=-\frac{\partial E_g(\phi)}{\partial \phi},
\end{eqnarray}
where $E_g(\phi)$ is the ground state state energy at flux $\phi$. \par
In the following we calculate the current
at several values of $W/t$ and $V/t$.
We mostly calculate the current in the diffusive regime.
In one dimension the localization length $\xi$ is given by
\begin{eqnarray}
\label{xi}
\xi &=& \frac{105at^2}{W^2} \quad (W \ll 2\pi t),\\
\xi &=& \frac{a}{\ln(W/2{\rm e}t)} \quad (W \gg 2\pi t),
\end{eqnarray}
where $a$ is the lattice constant \cite{kirk,weg,che}.
Equations above were derived
under the condition of the half-filled-band case
for the tight-binding model.
The system investigated here is not half-filled.
However, we use these equations to estimate the localization length,
since it is not far from half-filled.
We are mainly interested in the diffusive regime
and consider the value of $W$ such that $\xi>L=N_s a$. \par
%%%%%%%%%%%%%%%
%   Results   %
%%%%%%%%%%%%%%%
We first examine validity of the HFA.
For that purpose we calculate the persistent current
for systems of 4 electrons in 10 sites by the two methods:
the exact diagonalization and the HFA.
Average over site energy randomness is performed
over ten samples \cite{note}.
Flux dependence of the persistent current at $W=t$
and for $V=0$ to $20t$ is shown in Fig.1 for half period,
$0 \leq \phi \leq \phi_0/2$;
the current is an odd function of $\phi$,
and it is periodic with period $\phi_0$.
The current is normalized by the maximum persistent current
for clean ($W=0$) and non-interacting ($V=0$) system,
$I_0=(2et/N_s \hbar) \sin(N_e \pi/N_s) $\cite{che}.
In these systems the amplitude of the persistent current is suppressed
as the Coulomb interaction becomes larger.
The result by the exact diagonalization is in agreement with
that in Ref.\cite{ab}
where a system with 5 electrons in 10 sites is investigated.
Although the suppression is stronger,
we see the HFA gives qualitatively similar behavior.
The comparison is done at other values of $W$ also.
Figure 2 summarizes the results.
The vertical axis shows the magnitude of the current $I$
at $\phi=\phi_0/4$ divided by $I_{{\rm clean}}$
which is the value of the current
at $\phi=\phi_0/4$ in the clean, non-interacting system:
$I_{\rm {clean}}=I_0\sin(\pi/2N_s)/\sin(\pi/N_s) \simeq I_0/2$ \cite{che}.
The plot of non-interacting case (circle) is almost on the curve
represented by $\exp(-L/\xi(W))$.
The HFA results reproduce those
of the exact diagonalization qualitatively:
The persistent current is suppressed
by the Coulomb interaction in the diffusive regime, $W<3$.
It is remarkable that the slight enhancement of the current
in the localized regime ($W>3$) is also reproduced. \par
Now that we have found that the HFA gives
qualitatively correct behavior of the persistent current,
we investigate larger systems
where the exact diagonalization is impossible.
Specifically we consider systems with 40 electrons in 100 sites.
Figure 3 shows the flux dependence
of the persistent current at $W=0.25t$,
and Fig.4 shows $W$ dependence of the current.
In these figures average over 10 samples are shown.
For this 100-site systems the sample dependence is small
due to self-averaging.
Therefore 10 samples are enough.
These results also show
the suppression of the current with increase in $V$.
In this system $\xi \simeq L$ at $W \simeq t$,
so only the diffusive regime is considered.
The non-interacting case is also almost on the curve $\exp(-L/W(\xi))$.
Compared to the 10-site system it is seen that
the effect of $V$ is larger in the present larger system. \par
%%%%%%%%%%%%%%%%%%%
%   Discussions   %
%%%%%%%%%%%%%%%%%%%
 From the results stated above we can conclude that
the Coulomb interaction between electrons cause
the reduction of the persistent current
in the experimentally relevant diffusive regime.
The results found in Refs.\cite{ab} and \cite{monc}
are confirmed for larger systems.
In Ref.\cite{ab} it was suggested that
this suppression was attributed to the Mott-Hubbard transition.
If so, it should occur in a clean ($W=0$) system,
and in fact it was observed in Refs.\cite{ab} and \cite{monc}.
It is also seen in our results, Fig.2 (a).
On the other hand,
in the HFA the suppression of the current does not occur
in a clean sample at any filling and $V$,
since the eigenstates in the case of $V$=0 are still the eigenstates
and the ground state is also unchanged.
Nevertheless, in a dirty system,
the current is suppressed as $V$ increases
both in the case of the exact diagonalization and the HFA.
Furthermore the Coulomb effect is larger in disordered systems
in the diffusive regime.
This seems to teach us that the Mott-Hubbard transition is not essential
to the suppression of the current in a dirty system. \par
Our results are in disagreement with those in Ref.\cite{mg2}.
That may be considered due to the difference in models,
discrete or continuum, as some people \cite{some1,some2} pointed out.
However, in view of the fact that
the system with more lattice points approaches the continuum system,
our results that the Coulomb interaction suppress the current more
in the larger system cast some doubt
to the interpretation of the difference.
Another candidate for the reason of the enhancement in Ref. \cite{mg2}
would be truncation of the single electron states adopted there. \par
We cannot explain the experimental results
by our model and calculations.
It may be possible that
the enhancement of the current will occur
in multi-transverse channel systems.
These systems cannot be investigated by the exact diagonalization
because of the limitation in memories.
However,we can treat such a system by the HFA.
Results of such a calculation will be published in near future.\par
%%%%%%%%%%%%%%
%   Summary  %
%%%%%%%%%%%%%%
In summary we have first established the effectiveness of the HFA
for the present model.
With this approximation it has become possible
to investigate systems with more than 100 sites
even in the presence of the Coulomb interaction.
We have found that
the Coulomb interaction suppress the persistent current.\par
%%%%%%%%%%%%%%%%%%%%
%  Acknowledgment  %
%%%%%%%%%%%%%%%%%%%%
One of the authors (D. Y.) thanks Aspen Center for Physics
where part of this work was done.
This work was supported by a Grant-in Aid
for Scientific Research on Priority Area
^^ ^^ Computational Physics as a New Frontier in Condensed Matter Research"
(04231105) from the Ministry of Education, Science and Culture. \par
%%%%%%%%%%%%%%%%%%
%   References   %
%%%%%%%%%%%%%%%%%%

\newpage
%%%%%%%%%%%%%
%  Figures  %
%%%%%%%%%%%%%
\begin{center}
                FIGURE CAPTIONS\\
\end{center}
%
%-- Figure 1 --
%
FIG.1. Current $I/I_0$ vs flux $\phi/\phi_0$
for a 10-site 4-electron ring
at $W=t$, and
$V=0$ (circle), $5t$ (square), $10t$ (diamond), and $20t$ (triangle),
where $I_0=(2et/N_s \hbar) \sin(N_e \pi/N_s)$
is the maximum persistent current for clean, non-interacting system.
The panel (a) shows the results by the exact diagonalization
and panel (b) shows those by the HFA. \\
\\
%
%-- Figure 2--
%
FIG.2. Current $I/I_{{\rm clean}}$ vs randomness ($W$)
for a 10-site 4-electron ring
(a) by the exact diagonalization and
(b) by the HFA
for $V=0$ (circle), $5t$ (square), $10t$ (diamond) and $20t$ (triangle),
where
$I_{\rm {clean}}=I_0 \sin(\pi/2N_s)/\sin(\pi/N_s)$
is the value of the current at $\phi=\phi_0/4$
in the clean, non-interacting system.
The lines are  guides to the eye.\\
\\
%
%-- Figure 3 --
%
FIG.3. Current $I/I_0$ vs flux $\phi/\phi_0$
for a 100-site 40-electron ring
in the case of the HFA at $W=0.25t$,
and $V=0$ (circle), $t$ (square), $3t$ (diamond) and $5t$ (triangle).\\
\\
%
%-- Figure 4 --
FIG.4. Current $I/I_{{\rm clean}}$ vs randomness ($W$)
for a 100-site 40-electron ring
by the HFA
for $V=0$ (circle), $t$ (square), $3t$ (diamond) and $5t$ (triangle).
The lines are  guides to the eye.\\

\begin{thebibliography}{99}
\bibitem{note}Average over more samples (up to 500 samples)
has been done for typical choice of the parameters.
The behavior was qualitatively similar,
with slight deference in the amplitude.
\end{thebibliography}
\end{document}